\documentclass[twocolumn,showpacs,prd]{revtex4}
\usepackage{mathrsfs,bm,multirow}
\usepackage{longtable,lscape}
\usepackage{txfonts}
\usepackage{amssymb}
\usepackage{indentfirst}
\usepackage{graphicx,,booktabs}
\usepackage{color}
\usepackage{amssymb}

\usepackage[dvipdfm,  
            colorlinks, %
            citecolor=blue
            ]{hyperref}

\def\lrpartial{\buildrel\leftrightarrow\over\partial}

\begin{document}

\title{Anomalous radiative transitions between $h_b(nP)$ and $\eta_b(mS)$ and hadronic loop effect}
\author{Dian-Yong Chen$^{1,3}$}
\email{chendy@impcas.ac.cn}
\author{Xiang Liu$^{1,2}$\footnote{Corresponding author}}\email{xiangliu@lzu.edu.cn}
\author{Takayuki Matsuki$^4$}
\email{matsuki@tokyo-kasei.ac.jp}
\affiliation{$^1$Research Center for Hadron and CSR Physics,
Lanzhou University $\&$ Institute of Modern Physics of CAS,
Lanzhou 730000, China\\
$^2$School of Physical Science and Technology, Lanzhou University,
Lanzhou 730000, China\\
$^3$Nuclear Theory Group, Institute of Modern Physics, Chinese
Academy of Sciences, Lanzhou 730000, China\\
$^4$Tokyo Kasei University, 1-18-1 Kaga, Itabashi, Tokyo 173-8602,
Japan}

\begin{abstract}
In this work, we introduce the hadronic loop contribution to
explain the anomalous radiative transitions between $h_b(nP)$ and
$\eta_b(mS)$, which was recently observed by the Belle
Collaboration. Our calculation shows that the hadronic loop
mechanism associated with these known decay mechanisms can explain
why there exist anomalous radiative transitions between $h_b(nP)$
and $\eta_b(mS)$. This study deepens our understanding of the
decay mechanism of higher bottomonium radiative decays.

\pacs{13.20.Gd, 13.75.Lb}
\end{abstract}

\maketitle

\section{introduction}\label{sec1}

Besides reporting the evidence of a bottomonium $\eta_b(2S)$,
the Belle Collaboration recently measured several branching
ratios of the radiative transitions between $h_b(nP)$ ($n=1,2$) and
$\eta_b(mS)$ ($m=1,2$), i.e., 
$\mathcal{B}[h_b(1P)\to\eta_b(1S)\gamma]=(49.2\pm5.7^{+5.6}_{-3.3})\%$,
$\mathcal{B}[h_b(2P)\to\eta_b(1S)\gamma]=(22.3\pm3.8^{+3.1}_{-3.3})\%$
and $\mathcal{B}[h_b(2P) \to \eta_b(2S)\gamma]=
(47.5\pm10.5^{+6.8}_{-7.7})\%$ \cite{Mizuk:2012pb}. As indicated
in Ref. \cite{Mizuk:2012pb}, there exists a discrepancy between
these experimental values and theoretical expectations, where the
branching ratios \cite{Mizuk:2012pb} measured are a factor of
$1.2\sim 2.5$ higher than the corresponding theoretical results
given in Ref. \cite{Godfrey:2002rp}. This anomalous radiative
transition between $h_b(nP)$ and $\eta_b(mS)$ stimulates us to
propose a solution to this problem.

In the past years, experimentalists have observed many anomalous
hadronic decays and novel phenomena of higher charmonia and
bottomonia, which include the excessive non-$D\bar{D}$ component
of the inclusive $\psi(3770)$
\cite{Ablikim:2007zz,Ablikim:2006aj}, the Okubo-Zweig-Iizuka (OZI)
suppressed processes $\chi_{c1}\to\omega\omega,\phi\phi$ and
double-OZI suppressed $\chi_{c1}\to \omega\phi$
\cite{Ablikim:2011aa}, the $\chi_{cJ}$ radiative decays into a
light vector meson \cite{Bennett:2008aj,Ablikim:2011kv}, the
transitions of $\psi(4040)/\psi(4160)$ into $J/\psi\eta$
\cite{Wang:2012bg,Coan:2006rv,Ablikim:2012ht}, and anomalous large
rates of $e^+e^-\to \Upsilon(1S,2S)\pi^+\pi^-$ near the peak of
the $\Upsilon(5S)$ resonance \cite{Abe:2007tk}. Explaining these
phenomena, some theoretical efforts have been rewarded
\cite{Liu:2009dr,Zhang:2009kr,Chen:2009ah,
Liu:2009vv,Chen:2010re,Chen:2012nva,Meng:2007tk,Meng:2008dd},
where the hadronic loop effects as an important QCD
non-perturbative mechanism are introduced in these heavy
quarkonium decays. Successfully explaining the observed
experimental phenomena,
we realize that the
hadronic loop indeed plays a crucial role to the decays of heavy
quarkonia.

In addition, some discrepancies also appear between experimental
measurements and theoretical predictions on the radiative
transition between charmonia. The experimental measurements for
$J/\psi \to \eta_c \gamma$ and $\psi^\prime \to
\eta_c/\eta_c^\prime \gamma$ \cite{BESIII:2011ab} are much smaller
than the nonrelativistic and relativistic Godrey-Isgur quark model
predictions \cite{Godfrey:2002rp}. To alleviate the discrepancy
between experimental measurements and naive quark model
predictions, the meson loop contributions have been considered in
Refs. \cite{Li:2007xr,Li:2011ssa} and good agreements with the
experimental measurements have been archived. 

Along this line, in this work we adopt the hadronic loop effect to
investigate the anomalous radiative transitions between $h_b(nP)$
and $\eta_b(mS)$ observed by the Belle Collaboration
\cite{Mizuk:2012pb}. In this study, we first need to answer
whether the discrepancy between the experimental and theoretical
results of these radiative decays can be alleviated by including
the hadronic loop contribution added to the tree level diagram. If
this answer is affirmative, the hadronic loop effect, extensively
applied  to study intriguing higher charmonium and bottomonium
decays recently observed, can be further tested, which will make
our knowledge of non-perturbative QCD become abundant. In the next
section, the details of the hadronic loop effects on
$h_b(1P)\to\eta_b(1S)\gamma$, $h_b(2P)\to\eta_b(1S)\gamma$, and
$h_b(2P)\to\eta_b(2S)\gamma$ will be explicitly presented. For the
convenience of our presentation, these transitions are abbreviated
as $h_b(nP)\to \eta_b(mS)\gamma$ in the following sections.

This paper is organized as follows. After the introduction, we
illustrate the detailed formula of calculating the radiative
transitions between $h_b(nP)$ and $\eta_b(mS)$. In Sec.
\ref{sec3}, the numerical results are given by comparing them with
experimental data. The last section is devoted to a short summary.

\section{radiative transitions between $h_b(nP)$ and $\eta_b(mS)$}
\label{sec2}

In the naive quark model, the E1 transitions between $P$-wave and
$S$-wave spin-singlets are written as \cite{Godfrey:2002rp}
\begin{eqnarray}
\Gamma_{\text{QM}}(h_b(nP)\to \eta_b(mS)\gamma) =\frac{4}{9} \alpha
e_Q^2 \omega^3 \left|\langle {^1S_0} \left| r \right| {^1P_1}
\rangle\right|^2 \label{Eq:Red},
\end{eqnarray}
where $\alpha$ is the fine-structure constant and $e_Q=-1/3$
denotes the charge of the bottom quark in units of $|e|$. $\omega$
is the energy carried by the emitted photon. The spatial matrix
element $\langle {^1S_0} \left| r \right| {^1P_1} \rangle$ is
relevant to the radial wave functions of initial and final heavy
quarkonia.

\begin{figure}[htb]
\centering %
\scalebox{0.5}{\includegraphics{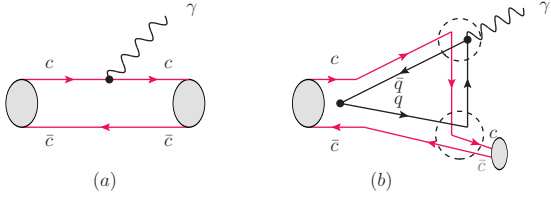}}
\caption{{The typical quark-level diagrams for the $h_b(nP)\to
\eta_b(mS)\gamma$ transition.}} \label{Fig:quark}
\end{figure}

{In Fig. \ref{Fig:quark}, we give the typical quark-level diagrams
depicting $h_b(nP)\to \eta_b(mS)\gamma$ in the naive quark model.
In the naive quark model, a loop diagram is usually ignored when
calculating the $h_b(nP)\to \eta_b(mS)\gamma$ transition. As
indicated by Belle \cite{Mizuk:2012pb}, the measured branching
ratio of $h_b(nP)\to \eta_b(mS)\gamma$ is not consistent with the
result obtained in the naive quark model \cite{Godfrey:2002rp}
which uses only Fig. \ref{Fig:quark} (a). Considering this
situation, we need to introduce the higher order diagram shown in
Fig. \ref{Fig:quark} (b), where $h_b(nP)$ first couples with two
virtual bottom and anti-bottom mesons, which then turn into
$\eta_{b}(mS)\gamma$ via exchanging bottom meson. However, it is
rather difficult to calculate this kind of diagram Fig.
\ref{Fig:quark} (b), so we would like to work in hadronic level
diagrams.} Here, we assume one to one correspondence between quark
tree diagrams and meson tree diagrams as well as quark loop
diagrams and meson loop diagrams. This is an assumption, but even
now it is clear indirectly. It would be worthwhile to test it in
future investigations when data will become more accurate.

{The hadronic tree level contribution to the E1 transition
corresponding to Eq. (\ref{Eq:Red}) is given as follows. Assuming
a point form factor of electromagnetic interaction and}
considering the gauge invariance of the photon field, we can
construct the Lorentz structure of the tree amplitude for the
$h_b(nP)\to \eta_b(mS)\gamma$ process as
\begin{eqnarray}
\mathcal{M}_{\text{QM}}= g_{\text{QM}} \epsilon_\gamma^{\mu}
\epsilon_{h_b}^\nu \left(g_{\mu \nu} -\frac{p_{\gamma \nu} p_{h_b
\mu}}{p_\gamma \cdot p_{h_b}}\right).\label{1}
\end{eqnarray}
Here the effective coupling $g_{\text{QM}}$ can be extracted by
comparing the decay widths, Eqs. (\ref{1}) and (\ref{Eq:Red}). The
sign of $g_ {\text{QM}}$ is set to be positive in the present
work. As indicated in the introduction, for $h_b(nP)\to
\eta_b(mS)\gamma$ transitions there exists a discrepancy between
the experimental measurement by Belle \cite{Mizuk:2012pb} and the
predictions by the naive quark model \cite{Godfrey:2002rp}.

\begin{figure}[htb]
\centering %
\scalebox{0.5}{\includegraphics{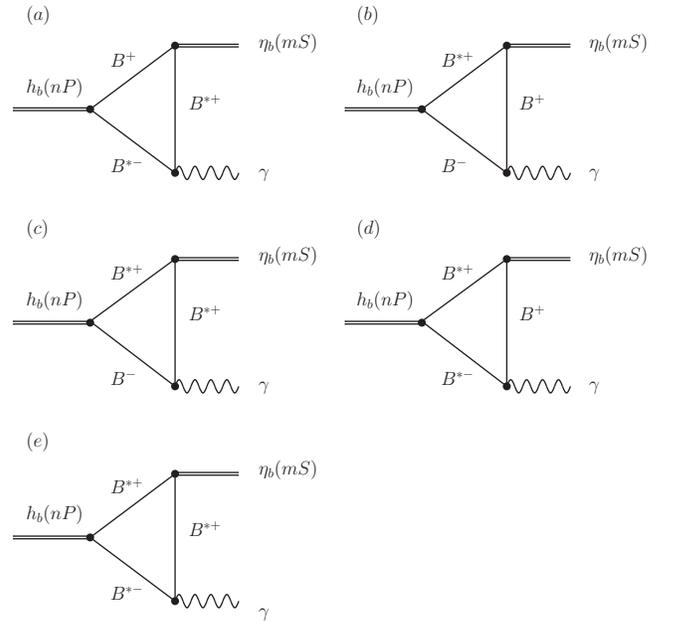}}
\caption{The hadronic triangle loop contributions to $h_b(nP) \to
\eta_b(mS) \gamma$. Replacing the charged bottom mesons with
neutral bottom mesons in diagrams (c) and (d), we can obtain the
remaining diagrams. \label{Fig:Tri}}
\end{figure}

In the following, we introduce the meson loop contributions other
than the tree contribution given by Eq. (\ref{1})
when studying $h_b(nP)\to \eta_b(mS)\gamma$ decays. Then we would
like to answer the question whether the discrepancy can be
alleviated by the hadronic loop effect or not. To calculate the
diagrams shown in Fig. \ref{Fig:Tri}, the effective Lagrangian
approach is adopted, where we use the following Lagrangians
constructed in heavy quark limit \cite{Cheng:1992xi, Yan:1992gz,
Wise:1992hn, Burdman:1992gh,Casalbuoni:1996pg}:
\begin{eqnarray}
\mathcal{L}_{h_b B^{(\ast)} B^{(\ast)}} &=& g_{h_b B^\ast B} h_b^\mu
(\bar{B}^\ast_{\mu} B + B^\ast_\mu \bar{B})+ ig_{h_b B^\ast B^\ast}
\varepsilon^{\mu \nu \alpha \beta}
\partial_{\mu} h_{b \nu} B^\ast_{\alpha} \bar{B}^\ast_{\beta},\nonumber\\
\mathcal{L}_{\eta_b B^{(\ast)} B^{(\ast)}} &=& i g_{\eta_b B^\ast B}
B^{\ast \mu}(\partial_\mu \eta_b \bar{B}-\eta_b
\partial_\mu  \bar{B}) +H.C. \nonumber\\& -& g_{\eta_b B^\ast B^\ast}
\varepsilon^{\mu\nu\alpha\beta}
\partial_\mu B^\ast_\nu \bar{B}^{\ast}_\alpha \partial_\beta\eta_b, \nonumber\\
\mathcal{L}_{B^{(\ast)} B^{(\ast)} \gamma} &=& i e A_{\mu} B^{-}
\lrpartial^{\mu} B^{+} + 
\Big\{\big( \frac{e}{4} g_{{B^{*+}B^-}\gamma} \varepsilon^{\mu \nu
\alpha \beta} F_{\mu \nu} \mathcal{B}^{*+}_{\alpha \beta} {B}^-
\nonumber\\& +&\frac{e}{4} g_{{B^{*0}{B}^0}\gamma} \varepsilon^{\mu
\nu \alpha \beta} F_{\mu \nu} \mathcal{B}^{*0}_{\alpha \beta}
\bar{{B}}^0 \big)+ h.c \Big\} +i e A_{\mu}  \nonumber\\ 
& \times &\left( g^{\alpha \beta} B_{\alpha}^{*-} \lrpartial^{\mu}
B_{\beta}^{*+}+ g^{\mu \beta} B_{\alpha}^{*-}
\partial^{\alpha} B_{\beta}^{*+} - g^{\mu \alpha }
\partial^{\beta} B_{\alpha}^{*-}  B_{\beta}^{*+}\right),\nonumber
\end{eqnarray}
where $\mathcal{B}_{\alpha\beta}^\ast=\partial_\alpha B_\beta^\ast
-\partial_\beta B_\alpha^\ast$. In the heavy quark limit, the
couplings among $h_b/\eta_b$ and the bottomed meson pair can be
related to two gauge coupling constants $g_1$ and $g_2$ by
\begin{eqnarray}
g_{h_b B^\ast B} &=& -2g_1 \sqrt{m_{h_b} m_B m_{B^\ast}} ,\nonumber\\
g_{h_b B^\ast B^\ast } &=&  2g_1 m_{B^\ast}/\sqrt{m_{h_b}},\nonumber \\
g_{\eta_b B^\ast B} &=& 2g_2 \sqrt{m_{\eta_b} m_B m_{B^\ast}},\nonumber\\
g_{\eta_b B^\ast B^\ast} &=& 2g_2 m_{B^\ast}/\sqrt{m_{\eta_b}},
\end{eqnarray}
where $h_b(nP)$ and $\eta_b(mS)$ are abbreviated as $h_b$ and $\eta_b$, respectively.
Thus, we need only to determine $g_1$ and $g_2$ corresponding to the couplings of $h_b(nP)$ and $\eta_b(mS)$, i.e.,
$g_1=-\sqrt{m_{\chi_{b0}(nP)}/3}/f_{\chi_{b0}(nP)}$ and $g_2=
\sqrt{m_{\Upsilon(mS)}}/ (2m_B f_{\Upsilon(mS)})$. The decay constants of
$P$-wave bottomonium  $f_{\chi_{b0}(1P)}$ and $f_{\chi_{b0}(2P)}$
can be related to the radial wave function of bottomonia
\cite{Lansberg:2009xh}, i.e., $f_{\chi_{b0}(1P)}=265$ MeV
and $f_{\chi_{b0}(2P)}=282$ MeV. In addition,
$f_{\Upsilon(1S)}=715$ MeV and $f_{\Upsilon(2S)}=482$ MeV are decay
constants of $\Upsilon(1S)$ and $\Upsilon(2S)$, respectively, which
are evaluated by their leptonic decay widths \cite{Beringer:1900zz}.
As for the $B^\ast B
\gamma$ coupling constant, we adopt $g_{B^{\ast +} B^+ \gamma }=
1.311\, \mathrm{GeV}^{-1}$ and $g_{B^{\ast 0} B \gamma} =-0.749
\,\mathrm{GeV}^{-1}$, both of which are estimated by the
light-front quark model \cite{Choi:2007se}.

Having these effective Lagrangians, we obtain the decay amplitudes
corresponding to the diagrams listed in Fig. \ref{Fig:Tri} for the
process $h_b(p_0)\to [B^{(*)}(p_1)B^{(*)}(p_2)]B^{(*)}(q) \to
\eta_b(p_3)\gamma(p_4)$, which read as
\begin{eqnarray}
\mathcal{M}_{a} &=& (i)^3 \int \frac{d^4q}{(2 \pi)^4} \Big[ g_{h_b
B^\ast B} \epsilon_{h_b \mu} \Big] \Big[i g_{\eta_b B^\ast B} (ip_{3
\rho}+ ip_{1\rho} )\Big] \nonumber\\&&\times\Big[ie
\epsilon_{\gamma}^\nu (g_{\alpha \beta} (-iq_\nu+ip_{2\nu})+ g_{\nu
\beta}(-iq_{\alpha}) -g_{\nu \alpha} (-ip_{2 \beta}) ) \Big]
\nonumber\\&&\times\frac{1}{p_1^2-m_B^2} \frac{-g^{\mu \alpha}
+p_2^\mu p_2^\alpha/m_{B^\ast}^2}{p_2^2 -m_{B^\ast}^2}
\frac{-g^{\rho \beta}+ q^\rho q^\beta/m_{B^\ast}^2}{q^2
-m_{B^\ast}^2}
\mathcal{F}(q^2), \nonumber\\
\\
\mathcal{M}_{b} &=& (i)^3 \int \frac{d^4q}{(2 \pi)^4} \Big[ g_{h_b
B^\ast B} \epsilon_{h_b \mu} \Big] \Big[ ig_{\eta_b B^\ast B}
(ip_{3\rho} -iq_{\rho}) \Big] \nonumber\\&&\times\Big[ie
\epsilon_{\gamma \nu} (-iq^\nu +ip_2^\nu) \Big] \frac{-g^{\mu \rho}
+ p_1^\mu p_1^\rho/m_{B^\ast}^2 }{p_1^2 -m_{B^\ast}^2}\frac{1}{p_2^2
-m_B^2} \nonumber\\&&\times\frac{1}{q^2 -m_B^2} \mathcal{F}(q^2),
\\
\mathcal{M}_{c} &=& (i)^3 \int \frac{d^4q}{(2 \pi)^4} \Big[ g_{h_b
B^\ast B} \epsilon_{h_b \mu} \Big] \Big[ -g_{\eta_b B^\ast B^\ast}
\varepsilon_{\theta \phi \alpha \beta} (-ip_1^\theta) (ip_3^\beta)
\Big] \nonumber\\&&\times\Big[\frac{e}{4} g_{B^\ast B \gamma}
\varepsilon_{\rho \lambda \delta \tau} \epsilon_{\gamma}^{\nu}
(ip_{4}^\rho g_{\nu}^\lambda -ip_{4}^\lambda g_{\nu}^\rho)
(-iq_{\delta}g_{\xi}^\tau \nonumber\\&&+ iq^{\tau} g_{\delta \xi} )
\Big] \frac{-g^{\mu \phi}+ p_1^\mu p_1^\phi/m_{B^\ast}^2 }{p_1^2
-m_{B^\ast}^2 } \frac{1}{p_2^2 -m_B^2}
\nonumber\\&&\times\frac{-g^{\alpha \xi} +q^\alpha
q^\xi/m_{B^\ast}^2}{q^2 -m_{B^\ast}^2} \mathcal{F}(q^2),
\\
\mathcal{M}_{d} &=& (i)^3 \int \frac{d^4q}{(2 \pi)^4} \Big[i g_{h_b
B^\ast B^\ast} \varepsilon_{\rho \mu \alpha \beta} (-ip_0^\rho)
\epsilon_{h_b}^\mu \Big]\nonumber\\&&\times \Big[ig_{\eta_b B^\ast
B} (ip_{3\lambda} -iq_{\lambda}) \Big] \Big[\frac{e}{4} g_{B^\ast B
\gamma} \varepsilon_{\theta\phi \delta \tau} \epsilon_{\gamma}^\nu
(ip_{4}^\theta g_{\nu}^\phi \nonumber\\&&-ip_{4}^\phi g_{\nu}^\theta)
(-ip_{2}^\delta g_{\xi}^\tau +ip_{2}^\tau g_{\xi}^\delta) \Big]
\frac{-g^{\alpha \lambda} + p_1^\alpha p_1^\lambda/m_{B^\ast}^2
}{p_1^2 -m_{B^\ast}^2 }\nonumber\\&&\times\frac{-g^{\beta \xi} +
p_2^\beta p_2^\xi/m_{B^\ast}^2 }{p_2^2 -m_{B^\ast}^2} \frac{1}{q^2
-m_B^2} \mathcal{F}(q^2),
\\
\mathcal{M}_{e} &=& (i)^3 \int \frac{d^4q}{(2 \pi)^4} \Big[i g_{h_b
B^\ast B^\ast} \varepsilon_{\rho \mu \alpha \beta} (-ip_0^\rho)
\epsilon_{h_b}^\mu \Big]\nonumber\\&&\times \Big[ -g_{\eta_b B^\ast
B^\ast} \varepsilon_{\theta \phi \delta \tau} (-ip_1^\theta)
(ip_3^\tau) \Big] \Big[ ie\epsilon_\gamma^\nu (g_{\lambda \xi}
(-iq_\nu +ip_{2\nu})  \nonumber\\ &&+g_{\nu \xi} (-iq_{\lambda})-
g_{\nu \lambda} (-ip_{2 \xi}))\Big] \frac{-g^{\alpha \phi} +
p_1^\alpha p_1^\phi/ m_{B^\ast}^2}{p_1^2 -m_{B^\ast}^2}
\nonumber\\&&\times\frac{-g^{\beta \lambda} + p_2^\beta p_2^\lambda/
m_{B^\ast}^2}{p_2^2 -m_{B^\ast}^2} \frac{-g^{\delta \xi} + q^\delta
q^\xi/ m_{B^\ast}^2}{q^2 -m_{B^\ast}^2} \mathcal{F}(q^2),
\end{eqnarray}
where the form factor $\mathcal{F}(q)$ is introduced to depict
the internal structures and off-shell effects of the exchanged
mesons as well as to remove the UV divergence in the loop integrals.
In the present work, the form factor can be parameterized as
\cite{Li:2011ssa}
\begin{eqnarray}
\mathcal{F}(q)= \prod_{i}
\frac{m_i^2-\Lambda_i^2}{q_i^2-\Lambda_i^2}
\end{eqnarray}
with $q_i=q,p_1, p_2$, where the parameter $\Lambda_i$ can be
parametrized as $\Lambda_i=m_i +\alpha \Lambda_{QCD}$ with $m_i$
denoting the corresponding intermediate bottom meson mass and
$\Lambda_{QCD}=220$ MeV. As illustrated in the caption of Fig.
\ref{Fig:Tri}, there remain two more diagrams, whose amplitudes
can be obtained by $\mathcal{M}_c$ and $\mathcal{M}_d$ by
replacing the corresponding masses and the coupling constants.

\begin{figure}[!h]
\centering %
\scalebox{0.5}{\includegraphics{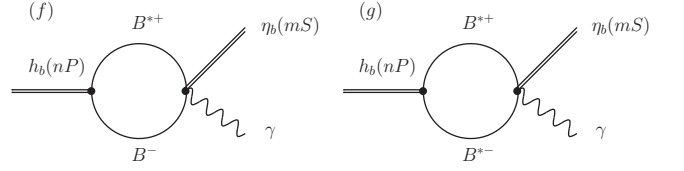}}
\caption{Contact diagrams for $h_b \to \eta_b \gamma$.
\label{Fig:Con}}
\end{figure}

In order to keep the gauge invariance of photon fields intact, the
contact diagrams given in Fig. \ref{Fig:Con} should also be
included in our calculation. The effective Lagrangian describing
vertex of $\eta_b(mS)$ interacting with $B^{*}B^{(*)}\gamma$ is
\begin{eqnarray}
\mathcal{L}_{\eta_b B^{\ast} B^{(\ast)}\gamma} &=&  g_{\eta_b
B^\ast B \gamma} B^{\ast \mu} \eta_b A_\mu B +H.C. \nonumber\\
&-& ig_{\eta_b B^\ast B^\ast \gamma} \varepsilon^{\mu \nu \alpha
\beta} A_\mu B^{\ast}_\nu \bar{B}^{\ast}_\alpha \partial_\beta
\eta_b \nonumber,
\end{eqnarray}
where the coupling constants $g_{\eta_b B^\ast B \gamma}$ and
$g_{\eta_b B^\ast B^\ast \gamma}$ can be obtained by the
corresponding ones of $\mathcal{L}_{\eta_b B^{(\ast)} B^{(\ast)}}$,
i.e., $g_{\eta_b B^\ast B \gamma} =e_{\bar{B}} g_{\eta_b B^\ast B}$
and $g_{\eta_b B^\ast B^\ast \gamma} =e_{B^\ast} g_{\eta_b B^\ast
B^\ast}$ with $e_{\bar{B}}(e_{B^\ast})$ denoting the charge of
$\bar{B}(B^{\ast})$ meson. By this effective Lagrangian, we obtain
the amplitudes corresponding to the contact diagrams shown in Fig.
\ref{Fig:Con}, which are
\begin{eqnarray}
\mathcal{M}_f &=& (i)^2 \int \frac{d^4q}{(2 \pi)^4} \Big[ g_{h_b
B^\ast B} \epsilon_{h_b}^\mu \Big] \Big[  g_{\eta_b B^\ast B \gamma}
\epsilon_\gamma^\nu \Big] \nonumber\\&&\times \frac{-g_{\mu \nu} +
q_{\mu}q_{\nu}/m_{B^\ast}^2 }{q^2-m_{B^\ast}^2} \frac{1}{(p_0-q)^2
-m_B^2} \mathcal{F}_{\mathrm{Con}}^2(q^2),\\ 
\mathcal{M}_g &=& (i)^2 \int \frac{d^4q}{(2 \pi)^4} \Big[ ig_{h_b
B^\ast B^\ast } \varepsilon_{\rho \mu \alpha \beta} (-ip_0^\rho)
\epsilon_{h_b}^\mu \Big] \nonumber\\ &&\times \Big[ -i g_{\eta_b
B^\ast B^\ast \gamma} \varepsilon_{\nu \lambda \theta \phi}
\epsilon_{\gamma}^\nu (ip_3^\phi) \Big] \frac{-g^{\alpha \lambda} +
q^{\alpha}q^{\lambda}/m_{B^\ast}^2 }{q^2-m_{B^\ast}^2} \nonumber\\
&&\times  \frac{-g^{\beta \theta} + (p_0^\beta
-q^\beta)(p_0^\theta-q^\theta)/m_{B^\ast}^2}{q^2 -m_{B^\ast}^2}
\mathcal{F}_{\mathrm{Con}}^2(q^2).
\end{eqnarray}
In the above amplitudes, another form factor
$\mathcal{F}_{\mathrm{Con}}(q^2)$ should be introduced as a
function of $q^2$, which cannot be arbitrary in order to keep the
gauge invariance.

By summing over all decay amplitudes of the triangle hadronic loop
diagrams and performing the loop integral, the general expression
of amplitude for the triangle diagrams is parameterized as
\begin{eqnarray}
\mathcal{M}_{\text{Tri}} 
=\epsilon_{\gamma}^{\mu} \epsilon_{h_b}^\nu \left(g_{\text{Tri}}
g_{\mu \nu}  - f_{\text{Tri}} \frac{p_{\gamma \nu} p_{h_b
\mu}}{p_{\gamma} \cdot p_{h_b}} \right).
\end{eqnarray}
On the other hand, the total amplitude of the contact diagram is
\begin{eqnarray}
\mathcal{M}_{\text{Con}} =\mathcal{M}_f+\mathcal{M}_g
=g_{\text{Con}} \epsilon_{\gamma}^{\mu} \epsilon_{h_b}^\nu g_{\mu
\nu},
\end{eqnarray}
where the detailed deduction of the amplitudes for two contact
diagrams is given in the Appendix.

With the above preparation, we finally get the total decay
amplitude of $h_b(nP)\to \eta_b(mS)\gamma$ considering only the
hadronic loop effect:
\begin{eqnarray}
\mathcal{M}_{\text{ML}} &=& \mathcal{M}_{\text{Tri}} +
\mathcal{M}_{\text{Con}} \nonumber\\
&=& \epsilon_{\gamma}^{\mu} \epsilon_{h_b}^\nu
\left[\left(g_{\text{Tri}}+g_{\text{Con}}\right)  g_{\mu \nu} -
f_{\text{Tri}} \frac{p_{\gamma \nu} p_{h_b \mu}}{p_{\gamma} \cdot
p_{h_b}} \right] \nonumber\\
&=&f_{\text{ML}}
\epsilon_{\gamma}^{\mu} \epsilon_{h_b}^\nu
\left( g_{\mu \nu}  -\frac{p_{\gamma \nu} p_{h_b
\mu}}{p_{\gamma} \cdot p_{h_b}} \right),\label{2}
\end{eqnarray}
which must be gauge invariant. In Eq. (\ref{2}), we define
$f_{\text{ML}}= g_{\text{Tri}}+g_{\text{Con}} = f_{\text{Tri}}$,
where $f_{\text{ML}}$ or $f_{\text{Tri}}$ is obtained by
calculating these triangle diagrams shown in Fig. \ref{Fig:Tri},
which is our main task in this work.

As indicated above, in this work we will test whether including the hadronic loop contributions
in $h_b(nP)\to \eta_b(mS)\gamma$ can alleviate a discrepancy between experimental and theoretical results.
The total amplitudes for $h_b(nP)\to \eta_b(mS)\gamma$ transition is composed of $\mathcal{M}_{\text{QM}}$ given by Eq. (\ref{1}) and $\mathcal{M}_{\text{ML}}$, i.e.,
\begin{eqnarray}
\mathcal{M}_{\text{tot}} &=&\mathcal{M}_{\text{QM}} + e^{i\phi}
\mathcal{M}_{\text{ML}} \nonumber\\&=& \left(g_{\text{QM}}+e^{i\phi}
f_{\text{ML}}\right) \epsilon_{\gamma}^{\mu} \epsilon_{h_b}^\nu
\left( g_{\mu \nu} -\frac{p_{\gamma \nu} p_{h_b \mu}}{p_{\gamma}
\cdot p_{h_b}} \right),
\end{eqnarray}
where $\phi$ is the phase angle between the amplitudes due to
different transition mechanisms. With these amplitudes, we can
estimate the transition width as
\begin{eqnarray}
\Gamma(h_b(nP)\to \eta_b(mS)\gamma) = \left| g_{\text{QM}}+e^{i\phi}
f_{\text{ML}}\right|^2 \frac{m_0^2-m_3^2}{24 \pi m_0^3 },
\end{eqnarray}
where $m_0$ and $m_3$ denote the masses of $h_b(nP)$ and $\eta_b(mS)$, respectively.

\section{Numerical results}\label{sec3}

\begin{table}[h!]
\caption{The $\gamma$ transition widths between $h_b(nP)\,
\{n=1,2\}$ and $\eta_b(mS)\, \{m=1,2\}$. The values of
$\Gamma_{\mathrm{QM}}$ are estimated in the naive quark model
\cite{Godfrey:2002rp}. $\Gamma_{\mathrm{ML}}$ denotes the
contributions from the meson loop, and their ranges are determined
by parameter $\alpha$. $\Gamma_{\mathrm{Tot}}$ are coherent
results including the amplitudes of both quark model and meson
loop contributions, which are dependent on parameter $\alpha$ and
phase angle $\phi$. The values of $\Gamma_{\mathrm{Exp}}$ are
estimated from the corresponding branching ratios from the Belle
Collaboration \cite{Mizuk:2012pb} together with some theoretical
results; the details are given in the text. \label{Tab:Width}}
\begin{tabular}{cccccc}
 \hline
Initial  & Final  & $\Gamma_{\mathrm{QM}}$ \cite{Godfrey:2002rp} &
$\Gamma_{\mathrm{ML}}$  &  $\Gamma_{\mathrm{Tot}} $  & $\Gamma_{\mathrm{Exp}}$ \\
states  & states  & keV & keV   & keV  & keV \\
 \hline
$h_b(1P)$   & $\eta_b(1S)$  & 37.0 & $0.1 \sim 6.4 $   & $12.6 \sim 74.2 $ & $38.9 \sim 70.0$  \\
$h_b(2P)$   & $\eta_b(1S)$  & 15.4 & $0.9 \sim 123.3$  & $0 \sim 203.5 $   & $18.9 \sim 123.1$\\
            & $\eta_b(2S)$  & 10.0 & $0.5 \sim 48.8 $  & $0 \sim 119.0 $   & $38.1 \sim 267.9$\\
 \hline
\end{tabular}
\end{table}

Besides the coupling constants mentioned in Sec. \ref{sec2}, the
masses involved in our calculation are $m_{h_b(1P)}=9899.1$ MeV,
$m_{h_b(2P)}= 10259.8 $ MeV, $m_{\eta_b(1S)}=9402.4$ MeV, and $
m_{\eta_b(1S)}=9999.0$ MeV \cite{Mizuk:2012pb}. The properties of
$P$ and $S$ wave singlet bottomonia are not well understood. At
present, the full widths of $h_b(1P)$ and $h_b(2P)$ are still not
yet measured by experiment. Thus, we have to adopt some
theoretical estimation, since we need to apply these resonance
parameters to obtain the experimental partial widths of
$h_b(nP)\to \eta_b(mS)\gamma$, where the Belle Collaboration gave
only the branching ratios of these decays \cite{Mizuk:2012pb}.
Besides the radiative transition to lower bottomonia, $h_b(1P)$
and $h_b(2P)$ dominantly annihilate into three gluons ($ggg$) or
two gluons plus one photon ($\gamma gg$). In Ref.
\cite{Godfrey:2002rp}, the widths of $h_b(1P)$ and $h_b(2P)$
decays into $ggg$ and  $\gamma gg$ were given, i.e.,
$\Gamma(h_b(1P)\to ggg)=50.8$ keV, $\Gamma(h_b(1P)\to \gamma
gg)=1.6$ keV, $\Gamma(h_b(2P)\to ggg)=50.2$ keV, and
$\Gamma(h_b(2P)\to gg\gamma)=1.6$ keV. Thus we can obtain the
branching ratio of $h_b(nP) \to \eta_b(mS)\gamma $, which is $
\mathcal{B}[h_b(nP) \to \eta_b(mS) \gamma ] =\Gamma_{h_b(nP) \to
\eta_b(mS) \gamma}/\Gamma^{\mathrm{Tot}}_{h_b(nP)} $, where
$\Gamma^{\mathrm{Tot}}_{h_b(nP)} =\sum_{m} \Gamma_{h_b(nP) \to
\eta_b(mS) \gamma} + \Gamma_{h_b(nP) \to ggg}+ \Gamma_{h_b(nP) \to
\gamma gg}$. With the partial decay widths of $h_b(nP) \to ggg
/\gamma gg$ from Ref. \cite{Godfrey:2002rp}, we can roughly obtain
the partial decay widths of $h_b(1P) \to \eta_b(1S)\gamma$,
$h_b(2P) \to \eta_b(1S)\gamma$ and $h_b(1P) \to \eta_b(2S)\gamma$
as $38.9 \sim 70.0$ keV, $18.9\sim 123.1$ keV, and $38.1 \sim
267.9$ keV, respectively.

The lack of experimental measurement of the decay width of
$h_n(1P)$ and $h_b(2P)$ leads to large uncertainties to the
partial decay width of $h_b(nP) \to \gamma \eta_b(mS)$. To compare
the coherent partial decay width including the amplitudes of quark
model and meson loop with those estimated from the branching
ratio, we adopt a large parameter space, which is $1\sim 4 $ for
$\alpha$ and $0 \sim 2 \pi$ for phase angle $\phi$. The $\gamma$
transition widths between $h_b(nP)\, \{n=1,2\}$ and $\eta_b(mS)\,
\{m=1,2\}$ are present in Table \ref{Tab:Width}. The results from
naive quark model are smaller than the lower limit of the
experimental measurements, especially for $h_b(2P)$ radiative
decay. The meson loop gives sizable contributions in the radiative
decay of $h_b(1P)/h_b(2P)$, and varies with parameter $\alpha$.
The discrepancies between experimental measurements and
theoretical estimation in the naive quark model can be alleviated.

\begin{figure}[htbp]
\centering %
\scalebox{0.85}{\includegraphics{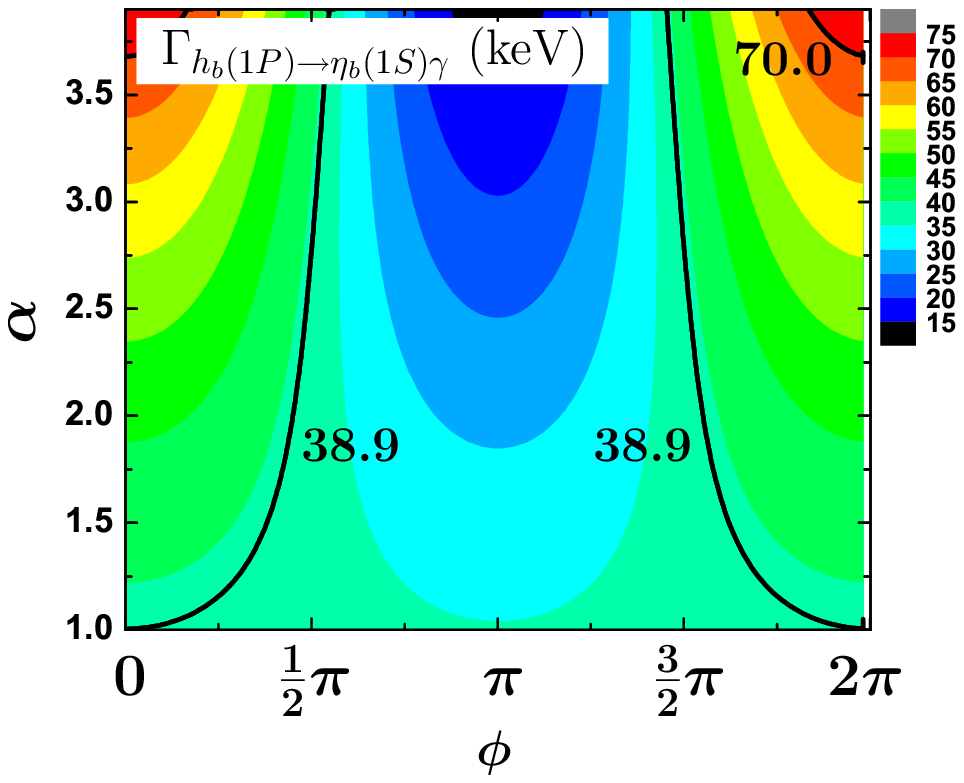}} %
\scalebox{0.85}{\includegraphics{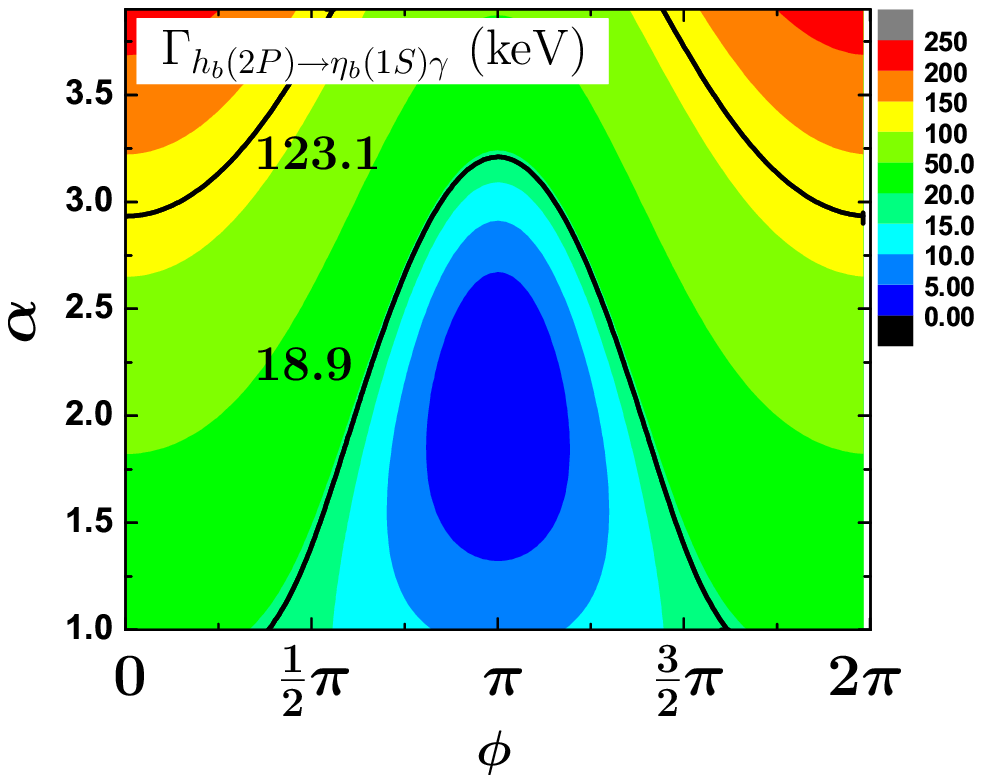}} %
\scalebox{0.85}{\includegraphics{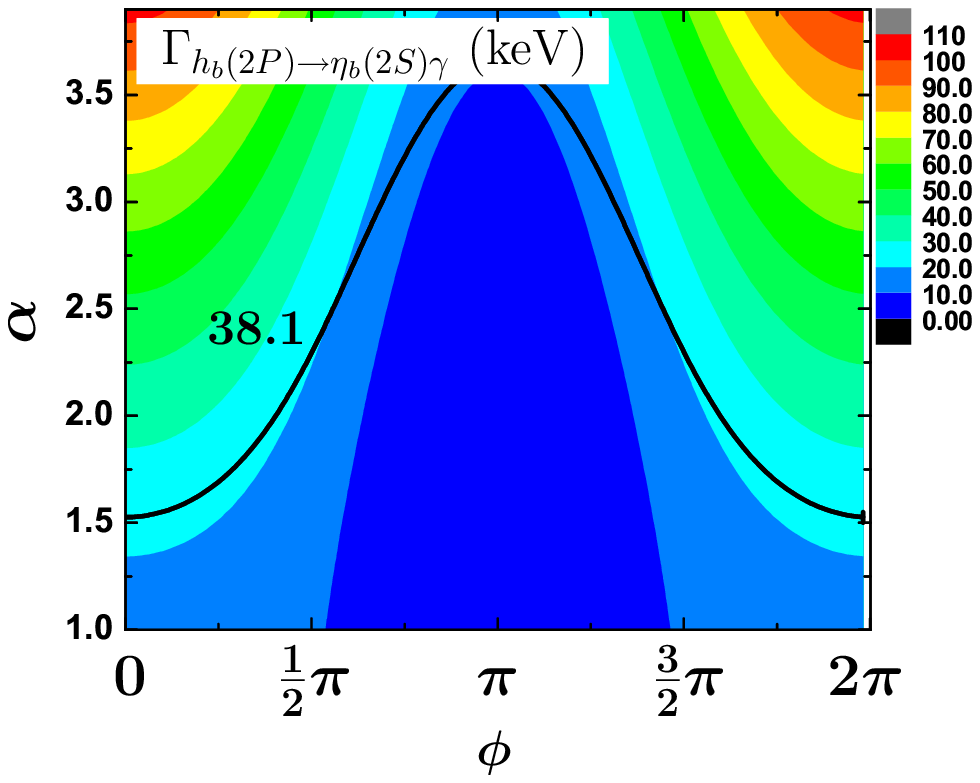}} %
\caption{(color online). The $\alpha$ and $\phi$ dependence of the
partial decay widths of $h_b(1P)\to\eta_b(1S)\gamma$,
$h_b(2P)\to\eta_b(1S)\gamma$, and $h_b(2P)\to\eta_b(2S)\gamma$
corresponding to the diagrams from top to bottom, respectively.
Here, the black curves are experimental lower and upper limits of
the corresponding decay width given by Belle \cite{Mizuk:2012pb}.
Units of values for decay widths are in keV.\label{Fig:nu}}
\end{figure}

In Fig. \ref{Fig:nu}, we show the decay widths of
$h_b(1P)\to\eta_b(1S)\gamma$, $h_b(2P)\to\eta_b(1S)\gamma$, and
$h_b(2P)\to\eta_b(2S)\gamma$ dependent on $\alpha$ in the
definition of $\Lambda_i$ and a phase factor $\phi$. The upper and
lower limits of the experimental measurements are also presented
as solid curves. The parts sandwiched by the curves are the
parameter spaces allowed by the experimental measurements. The
comparison between our numerical and the experimental results
indicates that the hadronic loop contributions to $h_b(nP)\to
\eta_b(mS)\gamma$ indeed can alleviate the discrepancy as
mentioned in Sec. \ref{sec1}. Thus, anomalous radiative
transitions between $h_b(nP)$ and $\eta_b(mS)$ can be understood
by introducing these hadronic loop diagrams. In addition, we also
notice that there exist common $\alpha$ and $\phi$ ranges where
the theoretical values are consistent with the experimental data
of $h_b(nP)\to \eta_b(mS)\gamma$ (see Fig. \ref{Fig:nu} for more
details). This phenomenon also reflects the similarity existing in
the $h_b(1P)\to\eta_b(1S)\gamma$, $h_b(2P)\to\eta_b(1S)\gamma$,
and $h_b(2P)\to\eta_b(2S)\gamma$ decays.

At present, there exists a large experimental range for these
discussed $h_b(nP)\to \eta_b(mS)\gamma$ transitions
\cite{Mizuk:2012pb}. We expect that future experiments can give a
more precise measurement, which will be useful to further
constrain our parameter range.

\section{short summary}\label{sec4}

Being stimulated by Belle's observation of anomalous radiative
transitions between $h_b(nP)$ and $\eta_b(mS)$, in this work we
study $h_b(nP)\to \eta_b(mS)\gamma$ transitions by introducing the
hadronic loop contributions. As shown in our numerical results,
the hadronic loop contributions play an important role to get
consistent results with the experimental data \cite{Mizuk:2012pb}.
These phenomena also show that the hadronic loop effects can be a
universal mechanism existing in the charmonium and botttomonium
hadronic and radiative decays, since there have been some
theoretical efforts \cite{Liu:2009dr,Zhang:2009kr,Chen:2009ah,
Liu:2009vv,Chen:2010re,Chen:2012nva,Meng:2007tk,Meng:2008dd} in
this direction.

With more and more accumulation of experimental data, some novel
phenomena of the decays of higher charmonium and bottomonium have
been revealed. If inclusion of the hadronic loops is a universal
non-perturbative QCD effect, we believe that the hadronic loop
mechanism should be further tested in the future by comparing the
results with more experimental observations, which is an
intriguing and fruitful research field.

\vfil

\section*{Acknowledgement}

This project is supported by the National Natural Science
Foundation of China under Grants No. 11222547, No. 11175073, No.
11005129, and No. 11035006, the Ministry of Education of China
(FANEDD under Grant No. 200924, SRFDP under Grant No.
20120211110002, NCET), the Fok Ying Tung Education Foundation (No.
131006), and the West Doctoral Project of Chinese Academy of
Sciences.

\appendix
\section{Contributions from Contact diagrams}\label{Sec:App-A}
As for the amplitude $\mathcal{M}_f$, one has
\begin{eqnarray}
\mathcal{M}_f &\propto& \epsilon_\gamma^\mu \epsilon_{h_b}^\nu \int
\frac{d^4 q}{(2 \pi)^4} \frac{-g_{\mu \nu} +q_{\mu}
q_{\nu}}{\left[q^2 -m_{B^\ast}^2 \right] \left[(p_0-q)^2
-m_B^2\right]} \mathcal{F}_{\text{Con}}(q^2) \nonumber\\
&=& \epsilon_\gamma^\mu \epsilon_{h_b}^\nu  \left(f_0 g_{\mu \nu}+
f_1 p_{0\mu} p_{0\nu}\right) =f_0 \epsilon_\gamma^\mu
\epsilon_{h_b}^\nu g_{\mu \nu}. \label{Eq:Con-f}
\end{eqnarray}
For the amplitude $\mathcal{M}_g$, it is in the form
\begin{eqnarray}
\mathcal{M}_g &\propto & \varepsilon_{\rho \mu \alpha \beta}
p_0^\rho \epsilon_{h_b}^\mu \varepsilon_{\nu \lambda \theta \phi}
\epsilon_{\gamma}^\nu p_3^\phi \int \frac{d^4q}{(2 \pi)^4}
\frac{-g^{\alpha \lambda} + q^{\alpha}q^{\lambda}/m_{B^\ast}^2
}{q^2-m_{B^\ast}^2} \nonumber\\ &\times& \frac{-g^{\beta \theta} +
(p_0^\beta -q^\beta)(p_0^\theta-q^\theta)/m_{B^\ast}^2}{(p_0-q)^2
-m_{B^\ast}^2} \mathcal{F}_{\mathrm{Con}}^2(q^2) \nonumber\\
&=& \varepsilon_{\rho \mu \alpha \beta} p_0^\rho \epsilon_{h_b}^\mu
\varepsilon_{\nu \lambda \theta \phi} \epsilon_{\gamma}^\nu p_3^\phi
\int \frac{d^4q}{(2 \pi)^4} \big(g^{\alpha \lambda} g^{\beta \theta}
+ (g^{\alpha \lambda} (q^\beta p_0^\theta \nonumber\\&-& q^\beta
q^\theta)-g^{\beta \theta} q^\alpha q^\lambda)+q^{\alpha} q^\lambda
q^\beta q^\theta - q^{\alpha} q^\lambda q^\beta
p_0^\theta\big)\nonumber\\&\times&1/\big([q^2 -m_{B^\ast}^2]
[(p_0-q)^2-m_B^2]\big)
\mathcal{F}_{\mathrm{Con}}^2(q^2) \nonumber\\
&=&  \varepsilon_{\rho \mu \alpha \beta} p_0^\rho \epsilon_{h_b}^\mu
\varepsilon_{\nu \lambda \theta \phi} \epsilon_{\gamma}^\nu p_3^\phi
\big( g_0 [gg]^{\alpha \lambda \theta \beta}+ g_1 [gp_0 p_0]^{\alpha
\lambda \theta \beta} \nonumber\\&+&g_2 [p_0p_0p_0p_0]^{\alpha
\lambda\theta \beta} \Big)\nonumber
\end{eqnarray}
where $[gg]^{\alpha \lambda \theta \phi}$ indicates the terms in
which these four Lorentz indices are shared with two metric
tensors, and $[gp_0 p_0]^{\alpha \lambda \theta \phi}$ and
$[p_0p_0p_0p_0]^{\alpha \lambda \theta \phi}$ are defined in the
same way. It is easy to find that the term proportional to $g_2$
vanishes after contracting all the Lorentz indices. The symbol
$[gp_0 p_0]^{\alpha \lambda \theta \phi}$ includes two different
cases. One is that $\alpha$ and $\beta$ are the Lorentz indices of
the metric tensor, i.e. $g^{\alpha \beta} p_0^\theta p_0^\phi$.
The other is that at least one of $\alpha$ and $\beta$ is the
Lorentz index of momentum $p_0$; i.e., these terms are
proportional to $p_0^\alpha$, $p_0^\beta$ or $p_0^\alpha
p_0^\beta$. After contracting all the Lorentz indices, one can
find that all these terms in two cases result to zero. Then for
$\mathcal{M}_g$ only the terms proportional to $g_0$ survive, that
is,
\begin{eqnarray}
\mathcal{M}_g&\propto &\varepsilon_{\rho \mu \alpha \beta} p_0^\rho
\epsilon_{h_b}^\mu \varepsilon_{\nu \lambda \theta \phi}
\epsilon_{\gamma}^\nu p_3^\phi [gg]^{\alpha \lambda \beta \theta}
\nonumber\\ &\propto& \varepsilon_{\rho \mu \alpha \beta} p_0^\rho
\epsilon_{h_b}^\mu \varepsilon_{\nu \lambda \theta \phi}
\epsilon_{\gamma}^\nu p_3^\phi g^{\alpha \lambda}g^{ \beta \theta}
\nonumber\\ &=& 2 p_0 \cdot p_3 \epsilon_{h_b}^\mu
\epsilon_{\gamma}^\nu g_{\mu \nu} -2 p_3 \cdot \epsilon_{h_b} p_0
\cdot \epsilon_{\gamma}.
\end{eqnarray}
In the initial state rest frame, $p_0^0=m_0$ and $\vec{p}_0=0$,
but for a real photon, the polarization vector has
$\epsilon_\gamma^0=0$ in both the Coulomb and axial gauge, and ,
thus, we have $p_0 \cdot \epsilon_\gamma=0$, and the amplitude
$\mathcal{M}_g$ should be in the form
\begin{eqnarray}
\mathcal{M}_g= g_0^\prime \epsilon_{h_b}^\mu \epsilon_{\gamma}^\nu
g_{\mu \nu}. \label{Eq:Con-g}
\end{eqnarray}
Then according to Eqs. (\ref{Eq:Con-f}) and (\ref{Eq:Con-g}), we
have
\begin{eqnarray}
\mathcal{M}_{\text{Con}}=\mathcal{M}_f+\mathcal{M}_g =g_{\text{Con}}
\epsilon_{h_b}^\mu \epsilon_{\gamma}^\nu g_{\mu \nu}.
\end{eqnarray}

\end{document}